\documentclass[review]{elsarticle}
\journal{Extreme Mechanics Letters}
\usepackage{graphicx}
\usepackage{float,array,caption,subcaption,multirow,stackengine,mathtools}
\biboptions{numbers,sort&compress}
\begin{document}

\begin{frontmatter}



\title{On Extracting Mechanical Properties from Nanoindentation at Temperatures up to 1000$^{\circ}$C}


\author[1]{James S. K.-L. Gibson\corref{cor1}}
\author[1]{Sebastian Schr\"{o}ders}
\author[1]{Christoffer Zehnder}
\author[1]{Sandra Korte-Kerzel\corref{cor2}}
\address[1]{RWTH Aachen University, Institut f\"{u}r Metallkunde und Metallphysik, Aachen, Germany}
\cortext[cor1]{{gibson@imm.rwth-aachen.de}}
\cortext[cor2]{{korte-kerzel@imm.rwth-aachen.de}}

\begin{abstract}
Alloyed MCrAlY bond coats, where M is usually cobalt and/or nickel, are essential parts of modern turbine blades, imparting environmental resistance while mediating thermal expansivity differences. Nanoindentation allows the determination of their properties without the complexities of traditional mechanical tests, but was not previously possible near turbine operating temperatures.

Here, we determine the hardness and modulus of CMSX-4 and an Amdry-386 bond coat by nanoindentation up to 1000$^{\circ}$C. Both materials exhibit a constant hardness until 400$^{\circ}$C followed by considerable softening, which in CMSX-4 is attributed to the multiple slip systems operating underneath a Berkovich indenter. 

The creep behaviour has been investigated via the nanoindentation hold segments. Above 700$^{\circ}$C, the observed creep exponents match the temperature-dependence of literature values in CMSX-4. In Amdry-386, nanoindentation produces creep exponents very close to literature data, implying high-temperature nanoindentation may be powerful in characterising these coatings and providing inputs for material, model and process optimisations. 
\end{abstract}


\end{frontmatter}

\section*{Introduction}
Gas turbines are widely used in both the aerospace industry and in land-based power generation. In both these sectors, there exists a strong driving force to increase operating temperatures for both fuel efficiency and a reduction in carbon emissions. In power generation, a one-percent improvement in efficiency can save \$20 million in fuel over the lifetime of the plant\cite{MRS:8606822}. In order to facilitate these increases, refinement of the coatings used in the turbine blades is required, as this prevents premature component failure from chemical attack and oxidation. Due to the small thickness of these coatings, traditional mechanical testing is extremely challenging. 

Nanoindentation -- whereby an indenter of a specific geometry is driven into the surface of a material under an applied load -- has been been widely used to determine the mechanical properties of thin films, either in the form of hard coatings\cite{Bushan2003125,Bull2005R393} or modified surface layers\cite{Kiener20122724,Armstrong2013251901}. The application of this technique and related nanomechanical testing at elevated temperatures is a relatively recent, but increasingly popular field, with indentation temperatures and publications rapidly increasing year-on-year\cite{Beake20022179,Sawant2008275,Trenkle2010817,Everitt20111221,Armstrong20151618,Wheeler2015354,Korte2012167,Hangen20151683,Huang2014162,Gibson2015380,Harris20161}.  It is also a potential breakthrough in the development of high temperature materials. As a technique, nanoindentation allows the quick determination of hardness, indentation modulus and creep resistance\cite{Li200211} on small volumes, enabling the mechanical testing of thin coatings to be carried out. Operation in vacuum additionally prevents deleterious oxidation of the sample or the indenter tip\cite{Korte2012167,Gibson2015380,Harris20161}. Nanoindentation has been applied to superalloy materials\cite{Durst2004312,Rehman2015202} and their bond coats\cite{Zotov20092064,Saeidi20111231,Girolamo20121989,Terberger2015360} with high-temperature nanoindentation recorded on CMSX-4 up to 400$^{\circ}$C\cite{Sawant2008275} and Ren\'e N4 up to 800$^{\circ}$C\cite{Li201665}. However, both works on high-temperature nanoindentation note the experimental difficulties - particular in the measurement of Young's modulus - due to creep during nanoindentation.

This work demonstrates the extension of high-temperature nanoindentation up to 1000$^{\circ}$C, allowing mechanical data, including creep exponents, to be determined at operationally-relevant temperatures on a sample of CMSX-4 and a 200 $\mu m$ thick Amdry-386 bond coat. In application, this bond coat provides environmental resistance and mediates thermal stresses between the superalloy and thermal barrier coating. The single crystalline CMSX-4 serves as a well-studied, but highly anisotropic reference material, while the polycrystalline bond coat is well suited to this and future studies by nanoindentation due to the more isotropic behaviour with small grain sizes, its intrinsically small size with regards to the thickness, and the interest in its high temperature properties for coating design.

\section*{Experimental Methods}
Single-crystal CMSX-4 (Ni-9.6Co-6.5Cr-6.5Ta-6.4W-5.6Al-0.6Mo-1Ti-3Re-0.1Hf, wt\%) was grown with an orientation close to [001] and subjected to a standard heat treatment to form the $\gamma - \gamma'$ structure. The sample was polished to a mirror finish using standard metallographic techniques, finished with 20 nm colloidal silica. An FEI Helios focussed ion beam (FIB) with EDAX EBSD (electron backscatter diffraction) and EDX (energy dispersive Xray) detectors was used to determine the grain orientations of the sample and to cross-section an indent for analysis (figure \ref{IndentCrossSection}). An Amdry-386 (Ni-22Co-17Cr-12Al-0.5Y-0.5Hf-0.2Si, wt\%) bond coat was thermally-sprayed onto the superalloy. EBSD performed on the superalloy and bond coat (see supplementary material) showed the Amdry-386 consists mainly of 2-3 $\mu m$, equiaxed grains.

Tests were performed using a MicroMaterials NanoTest P-3 system modified by a custom-built vacuum chamber ($\sim$$10^{-5}$ mbar). Indents were performed to a depth of 1 $\mu m$ using a an independently-heated Berkovich sapphire tip at a constant load rate of 2 mN/s, followed by a 30 second dwell period and 5 mN/s unloading rate. Up to 700$^{\circ}$C, at least 18 indents were made in the CMSX-4 and 12 indents in the bond coat at each temperature. At 900$^{\circ}$C and 1000$^{\circ}$C only two indents were made in each of the CMSX-4 and Amdry-386 as a larger number of indents at 1000$^{\circ}$C resulted in blunting of the tip due to tip-sample reactions and/or deposition of sample material as confirmed by atomic force microscopy (AFM) (see supplementary material). The diamond area function (DAF) of the tip was determined prior to testing and after testing at 700$^{\circ}$C by indenting into fused silica at room temperature. Additional indents to 500 mN in fused silica were used to calibrate the stiffness of the load frame. The initial DAF was used to analyse the data\cite{Oliver19921564} up to 700$^{\circ}$C and the post-700$^{\circ}$C DAF - which was only minimally different - was used for the data obtained at 900$^{\circ}$C and 1000$^{\circ}$C. Due to the extreme temperatures used, the temperature-dependent Poisson's ratios ($\nu$) of CMSX-4 and Al$_2$O$_3$ and temperature-dependent Young's modulus ($E$) of Al$_2$O$_3$ were used for the conversion of reduced modulus to Young's modulus for the CMSX-4 and Amdry-386 ($\nu$ from \cite{Sieborger200126} and \cite{Soga1966355}, $E$ from \cite{WACHTMAN1959254}). This temperature-dependence corresponds to a change of $\sim$8\% in $E_{Al_2O_3}$ and $\nu_{CMSX-4}$, and $\sim$5\% in $\nu_{Al_2O_3}$.

\section*{Results}
\subsection*{Nanoindentation Data Analysis}
No variation of hardness with position was observed in either material at a given temperature. This is unsurprising for single-crystal CMSX-4 and the small grain size of the bond coat relative to the indent size\cite{Chen20064289}. The crystal structure of the bond coat was not explicitly determined, but the successful indexing by EBSD with a nickel input file, as well as literature data\cite{Rathod2014432}, strongly suggests an FCC structure is present. No plastic anisotropy is therefore expected. As a result, data from each material were averaged for better statistics.

\begin{figure}[ht!]
\centering
\begin{subfigure}{.5\textwidth}
  \centering
  \includegraphics[width=.9\linewidth]{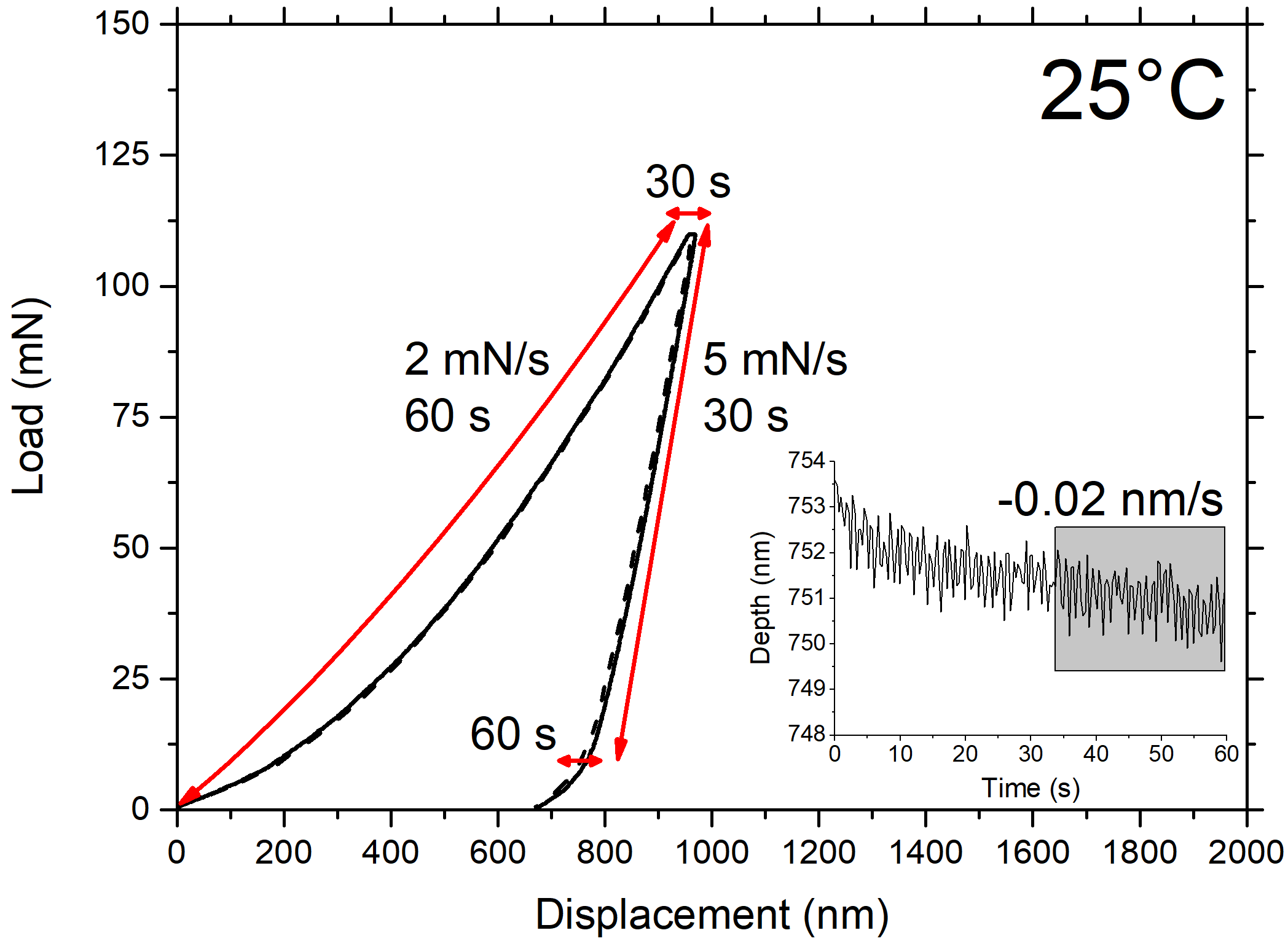}
  \caption{}
  \label{Hardness1}
\end{subfigure}%
\begin{subfigure}{.5\textwidth}
  \centering
  \includegraphics[width=.9\linewidth]{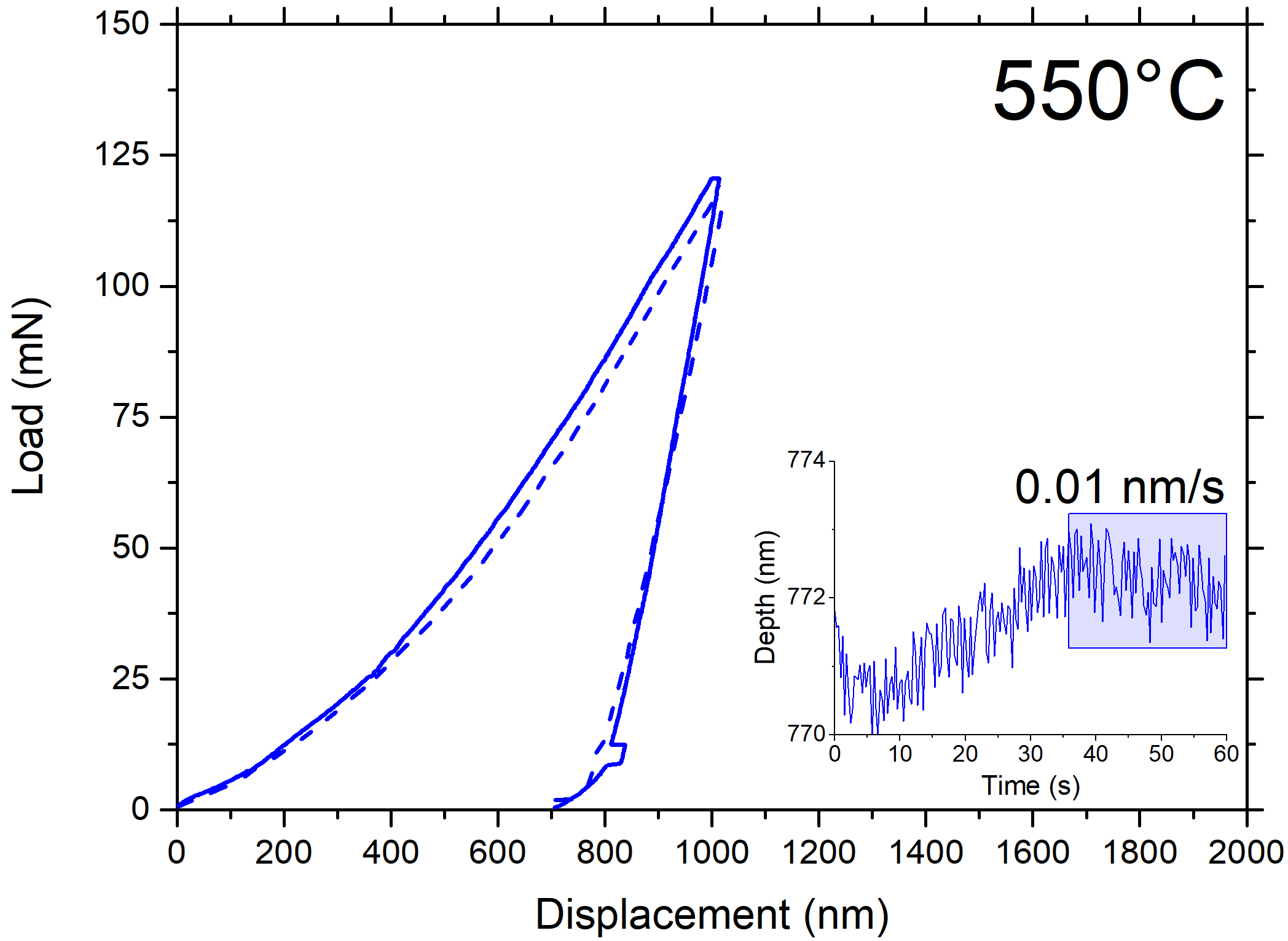}\\
  \caption{}
  \label{Hardness2}
\end{subfigure}
\begin{subfigure}{.5\textwidth}
  \centering
  \includegraphics[width=.9\linewidth]{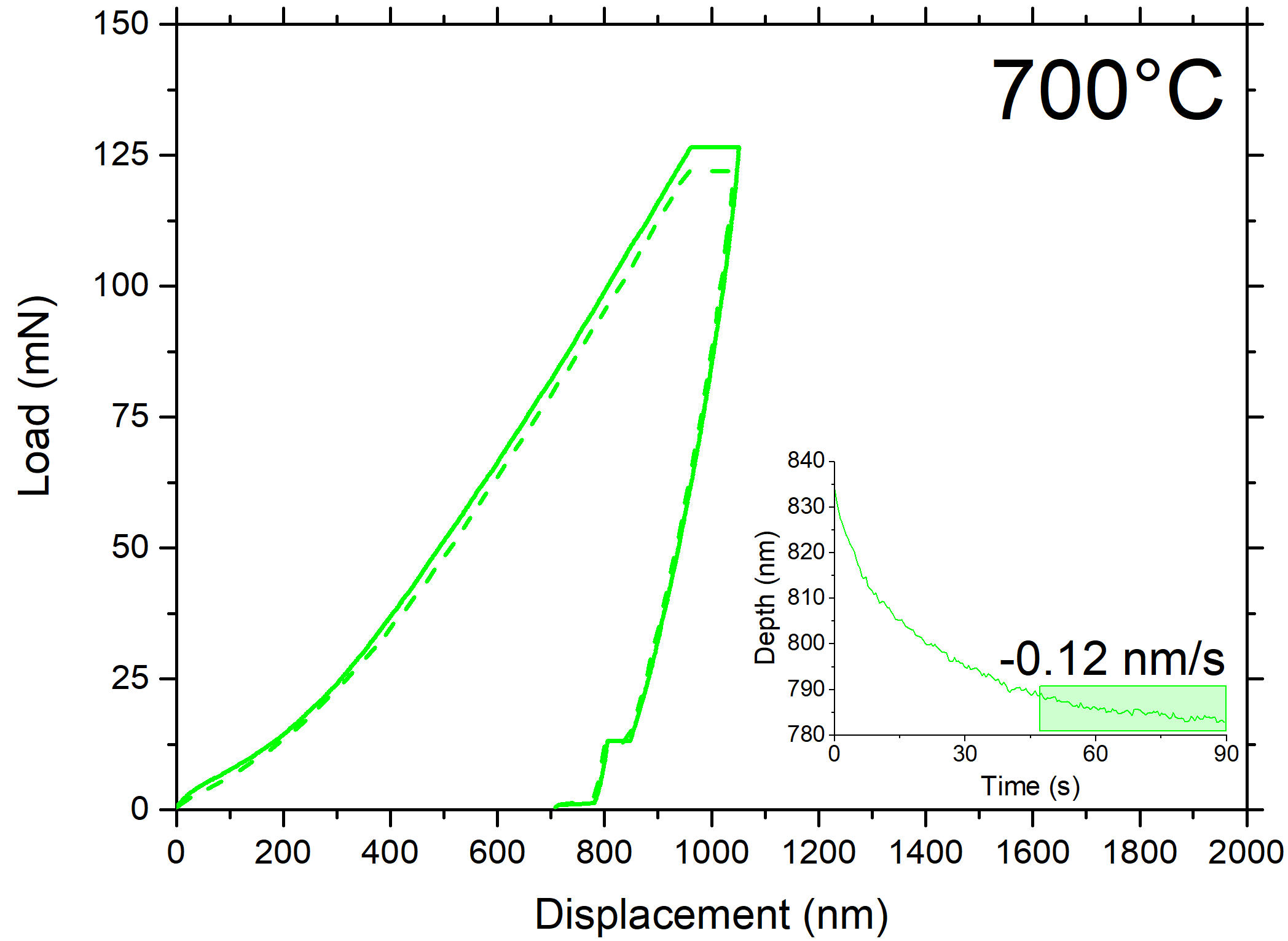}\\
  \caption{}
  \label{Hardness3}
\end{subfigure}%
\begin{subfigure}{.5\textwidth}
  \centering
  \includegraphics[width=.9\linewidth]{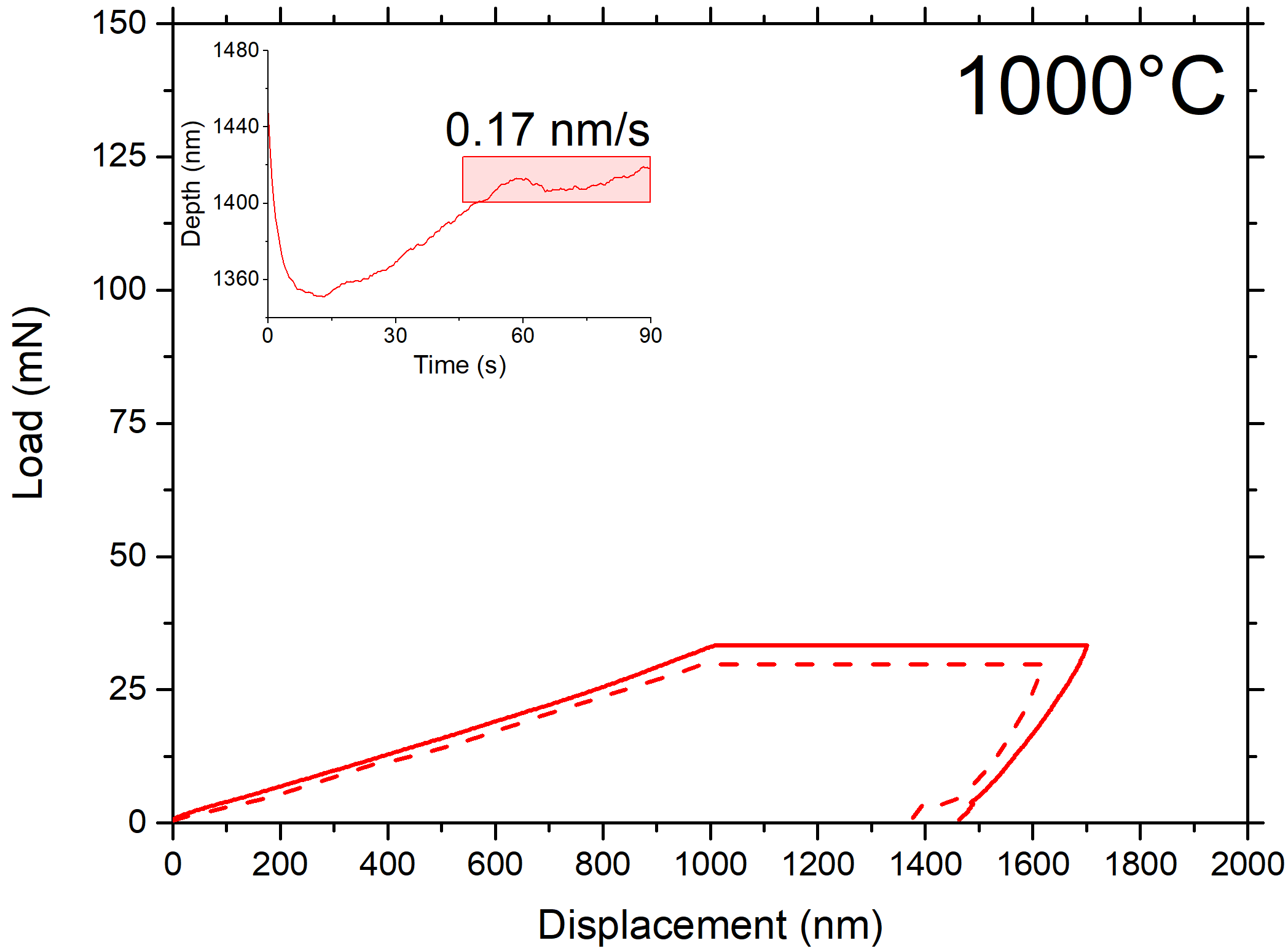}\\
  \caption{}
  \label{Hardness4}
\end{subfigure}
\caption{An individual load-displacement data at 25$^{\circ}$C (a), 550$^{\circ}$C (b), 700$^{\circ}$C (c) and 1000$^{\circ}$C (d) from Amdry-386 and CMSX-4 (Amdry data shown dashed). For the data at 25$^{\circ}$C, the labels show the times for each indentation segment. (N.B. the rates were kept constant, so that the loading and unloading segments change accordingly, while the hold segments remain the same.)\\ Inset: the drift hold segments, measured at 90\% unload at the end of indentation. Note the significant change in Y axis at 700$^{\circ}$C (c) and above (d) due to the influence of creep on the unloading data. The measurement of thermal drift is carried out over the last 40\% of the curve, highlighted. The significant, negative change in depth before this is the influence of creep.}
\label{LoadDispTemps}
\end{figure}

Load-displacement behaviour with temperature is shown in figure \ref{LoadDispTemps}, with all data plotted with the same scale and the time of each segment (at 25$^{\circ}$C) given in figure \ref{Hardness1}. Thermal drift is measured at 90\% unload, where the load-displacement data imply that above 550$^{\circ}$C, thermal drift becomes significantly worse. There is, however, a strong influence of creep on these data. As shown inset -- particularly in the 700$^{\circ}$C and 1000$^{\circ}$C curves -- there is a significant amount of initial unloading due to (reverse) creep, before the measured displacement is only from thermal drift, analogous to the viscoelastic effects noted in the ASTM ISO14577-1 standard for nanoindentation\cite{ASTMNano}.

\begin{figure}[ht!]
\centering
\begin{subfigure}{.5\textwidth}
  \centering
  \includegraphics[width=.9\linewidth]{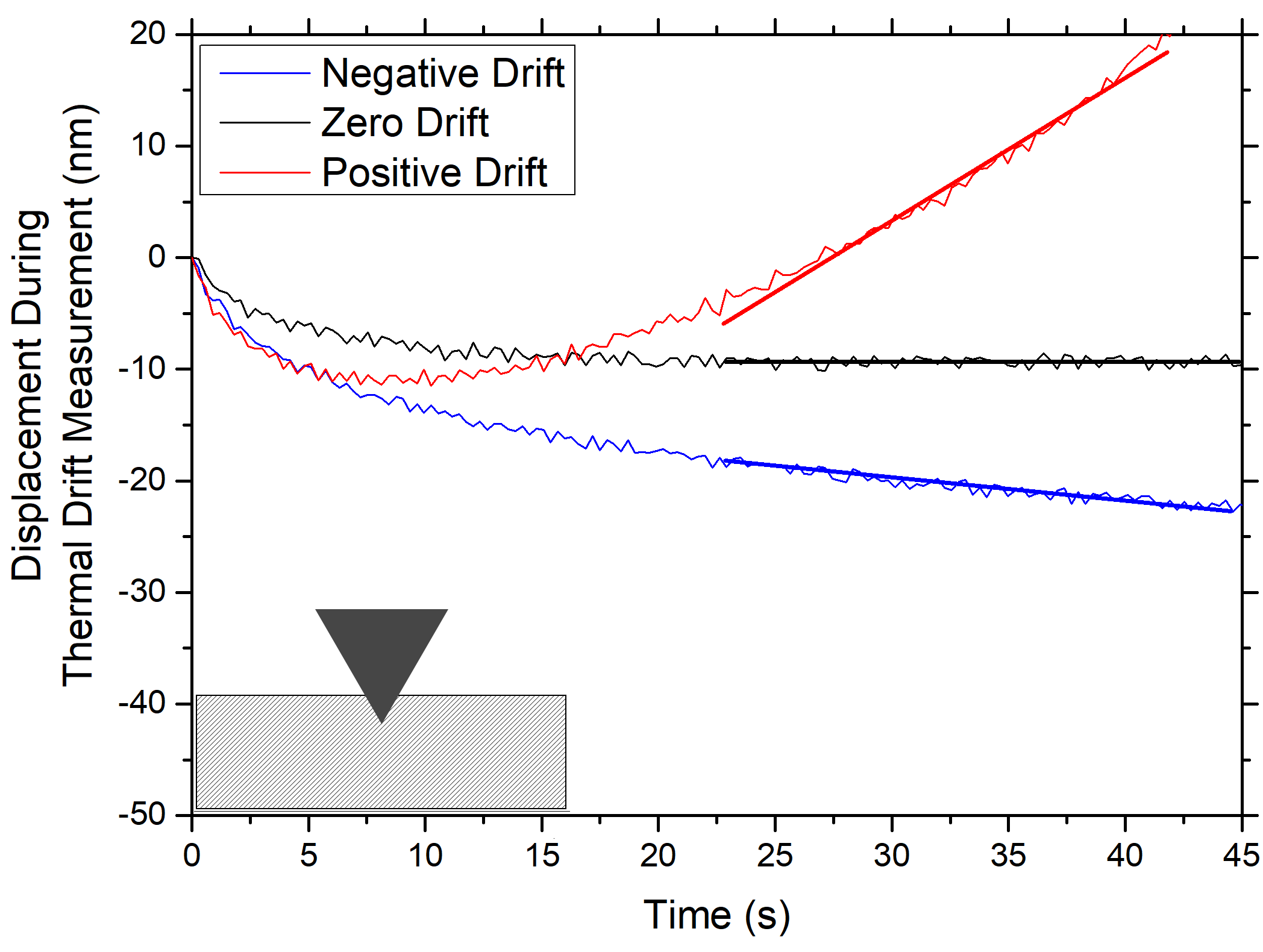}
  \caption{}
  \label{CreepThermalDriftIndent}
\end{subfigure}%
\begin{subfigure}{.5\textwidth}
  \centering
  \includegraphics[width=.9\linewidth]{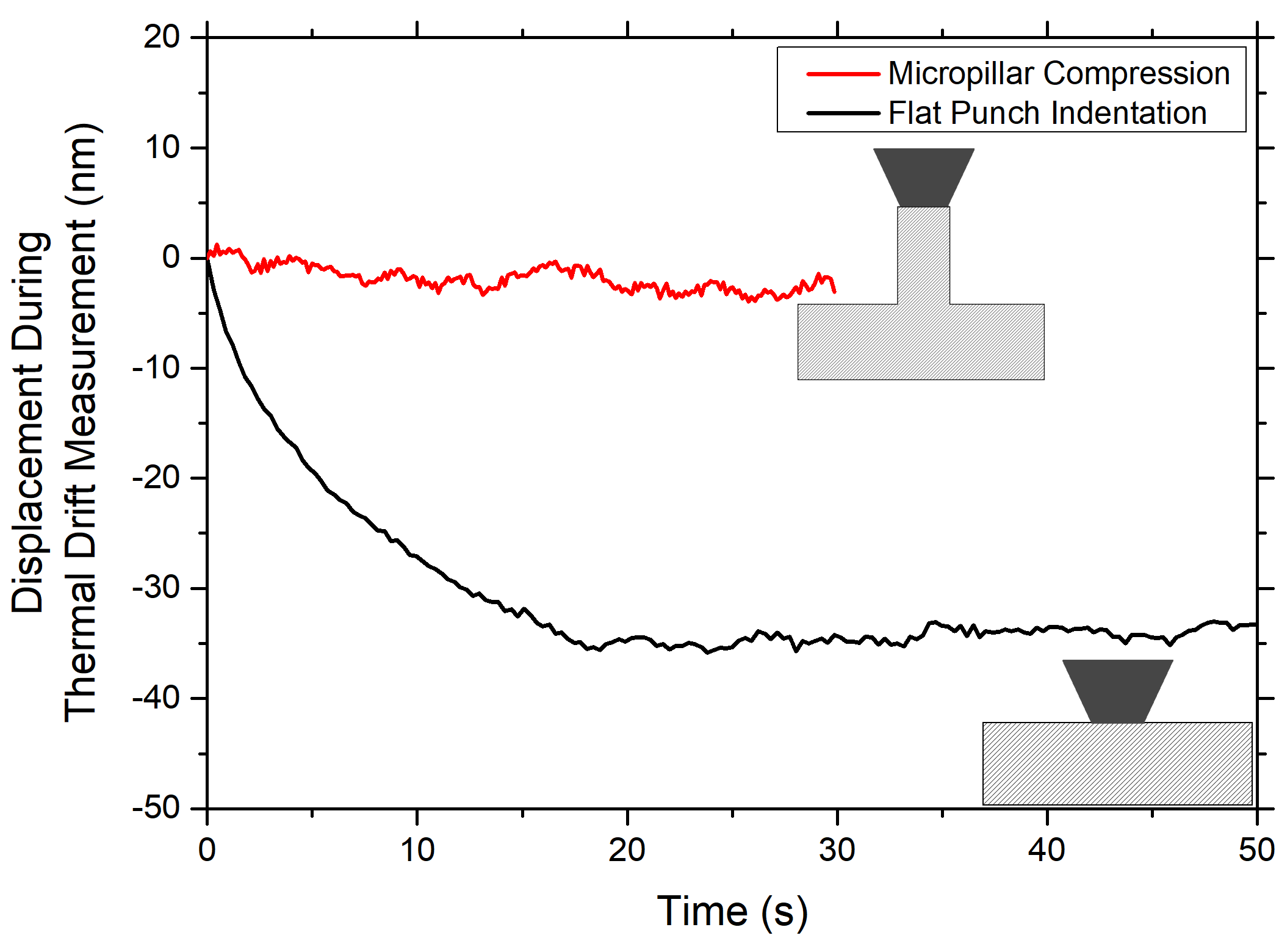}\\
  \caption{}
  \label{PillarVsPunch}
\end{subfigure}
\caption{(\textbf{a}) Three indents made into the sample (confining pressure) using a Berkovich tip during tuning at high temperature. (\textbf{b}) A comparison of the post-indentation thermal drift measurement for a micropillar (no confining pressure) and an indent made locally to the pillar at the same temperature (confining pressure). In all cases, the confining pressure drives reverse creep causing the initial rapid unloading, before a near-constant displacement rate is seen due to the thermal (in)stability. Note even when drift in the opposite direction is seen due to thermal mismatch between the tip and sample, this reverse creep remains. The micropillar data was collected as part of a previous publication\cite{Korte2012167} and re-analysed here.}
\label{PillarCreep}
\end{figure}

The driving force for this reverse movement of the indenter is given by the confined material around the tip. The effect of stress-state is shown directly by a comparison of drift segments on CMSX-4 at 630$^{\circ}$C on a micropillar, i.e. without confining pressure driving reverse creep (for further experimental details, see \cite{Korte2012167}, from where this data is taken and re-analysed), and away from the pillar using the same temperature settings and flat punch, producing a three-dimensional stress field similar to pyramidal indentation (figure \ref{CreepThermalDriftIndent}). Indents made using a Berkovich tip are shown, with the indenter at three different temperatures during tuning of the indentation contact, resulting in negative, zero and positive thermal drift rates. While a near-constant depth is maintained in the thermally stable pillar compression, the drift segment of the flat punch shows the same displacement out of the sample as observed in nanoindentation, before the depth reaches a near-constant value. The drift data from the Berkovich indentations demonstrate that this reverse movement is not due to thermal mismatch: it is present in all three drift segments before the true mismatch drift is measured in the system. This demonstrates the necessity of a `stabilisation period' before drift rates are measured and drift corrections are applied. When measured over the last 40\% of the indentation displacement-time data - the highlighted region of the insets in figure \ref{LoadDispTemps} - a thermal drift rate of -0.12 nm/s and +0.17 nm/s is measured at 700$^{\circ}$C and 1000$^{\circ}$C, respectively.

\subsection*{Mechanical Data}
\begin{figure}[ht!]
\centering
\begin{subfigure}{.5\textwidth}
  \centering
  \includegraphics[width=.9\linewidth]{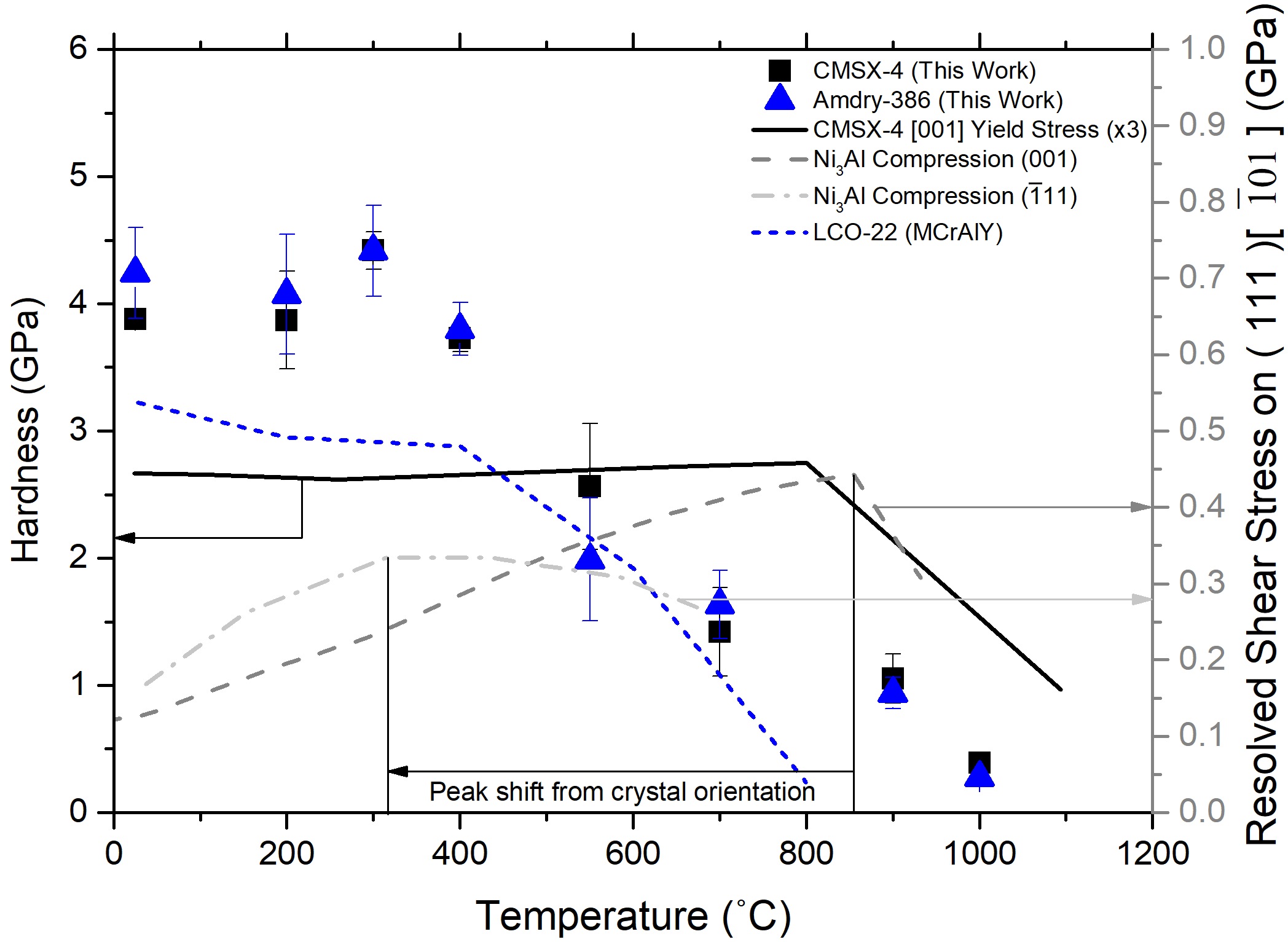}
  \caption{}
  \label{AllHardness}
\end{subfigure}%
\begin{subfigure}{.5\textwidth}
  \centering
  \includegraphics[width=.9\linewidth]{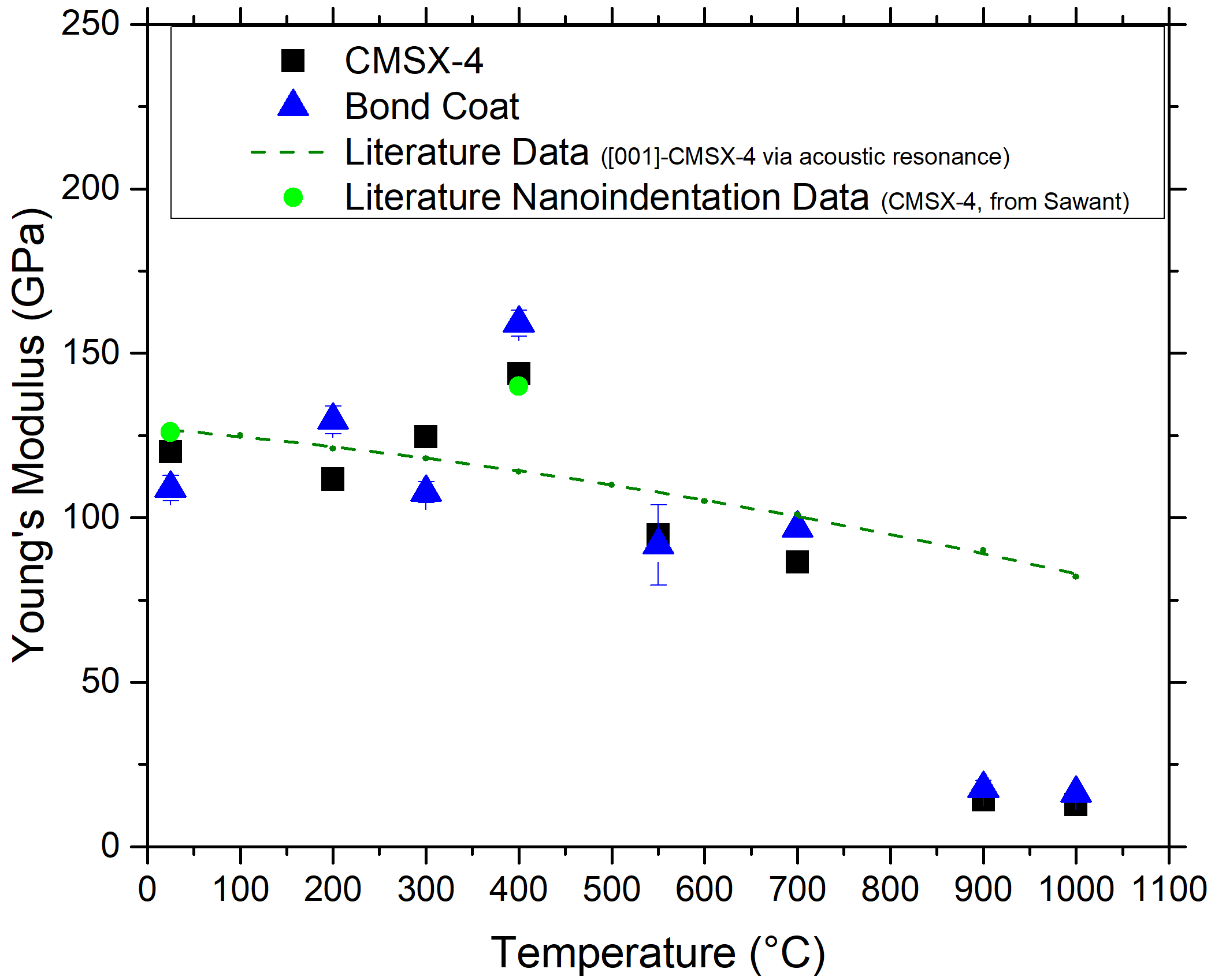}\\
  \caption{}
  \label{AllModulus}
\end{subfigure}
\caption{(\textbf{a}) Hardness with temperature of CMSX-4 and Amdry-386. A constant hardness is seen up to 400$^{\circ}$C. The error bars given are one standard deviation of the scatter in hardness. A comparison of nanoindentation hardness with literature data is given: three times the yield stress of [001]-oriented CMSX-4\cite{Sengupta199473} and resolved shear stresses under compression of the 0.2\% off-set yield stress of two single-crystal samples of Ni$_3$Al\cite{Ezz1982921}. To compare with the Amdry-386, the conventionally-measured hardness of LCO-22\cite{Taylor19969} is also shown.(\textbf{b}) Young's Modulus with temperature of CMSX-4 and Amdry-386. The error bars are one standard deviation of the scatter in modulus. The line shows the modulus of [001]-oriented CMSX-4 with temperature as found in the literature\cite{Sieborger200126} and values of Young's modulus measured by nanoindentation\cite{Sawant2008275}. Neither the data from this work nor the data from Sawant\cite{Sawant2008275} have been corrected for creep.}
\label{Hardness}
\end{figure}

The hardness-temperature behaviour of the sample is shown in figure \ref{AllHardness}. The hardness of the materials was measured at 4 GPa at room temperature, matching well with other nanoindentation tests\cite{Sawant2008275}. This hardness is constant until 400$^{\circ}$C, before steadily falling to 0.5 GPa at 1000$^{\circ}$C.  The indentation modulus as determined by an Oliver-Pharr fit\cite{Oliver19921564} is shown in figure \ref{AllModulus}. There are small deviations from the literature values from acoustic resonance, the largest at 400$^{\circ}$C. However, nanoindentation elsewhere\cite{Sawant2008275} measured near-identical values before the authors applied a creep correction. As little creep was measured here at these temperatures, we did not apply such a correction, and the data are overall consistent with the literature up to 700$^{\circ}$C. At 900$^{\circ}$C and 1000$^{\circ}$C the measured moduli are extremely low, due to an improper Oliver-Pharr fit to the unloading data, namely the assumption that the unloading is purely elastic, where in fact there is still significant time-dependent deformation occurring. The creep rates at these temperatures are extremely high (see figure \ref{StressExponent}) and this time-dependent deformation is still visibly present in the unloading data (figure \ref{Hardness4}). 

\begin{figure}[ht!]
\centering
\begin{subfigure}{.5\textwidth}
  \centering
  \includegraphics[width=.9\linewidth]{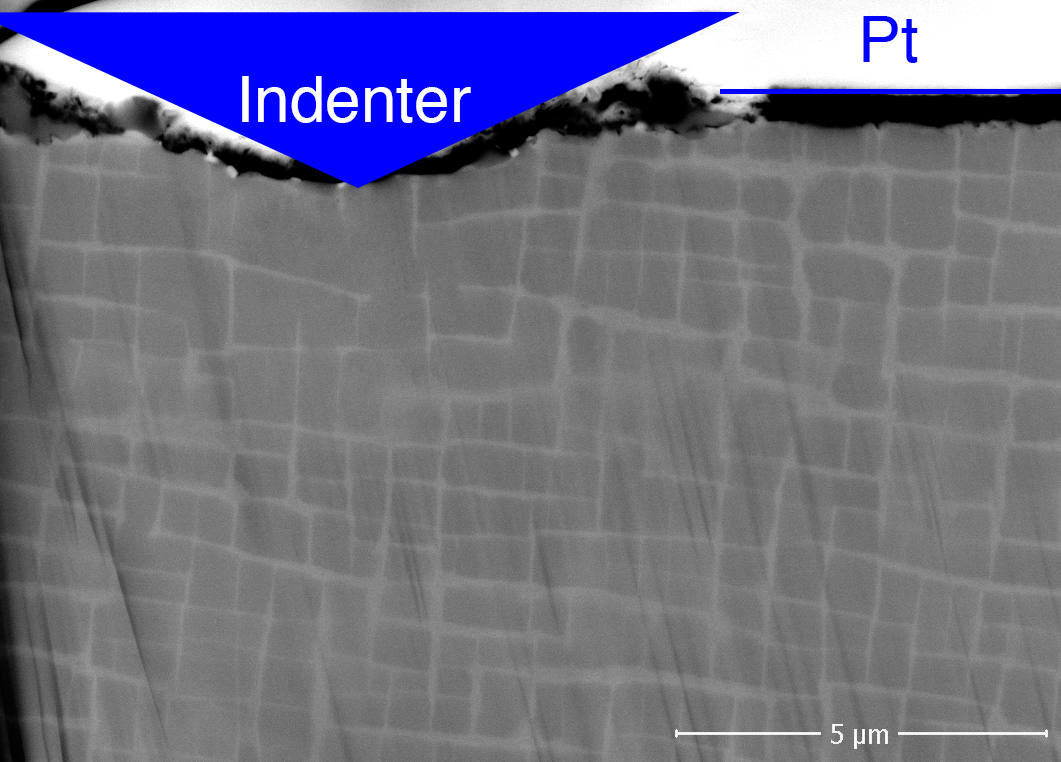}
  \caption{}
  \label{IndentCrossSection1}
\end{subfigure}%
\begin{subfigure}{.5\textwidth}
  \centering
  \includegraphics[width=.9\linewidth]{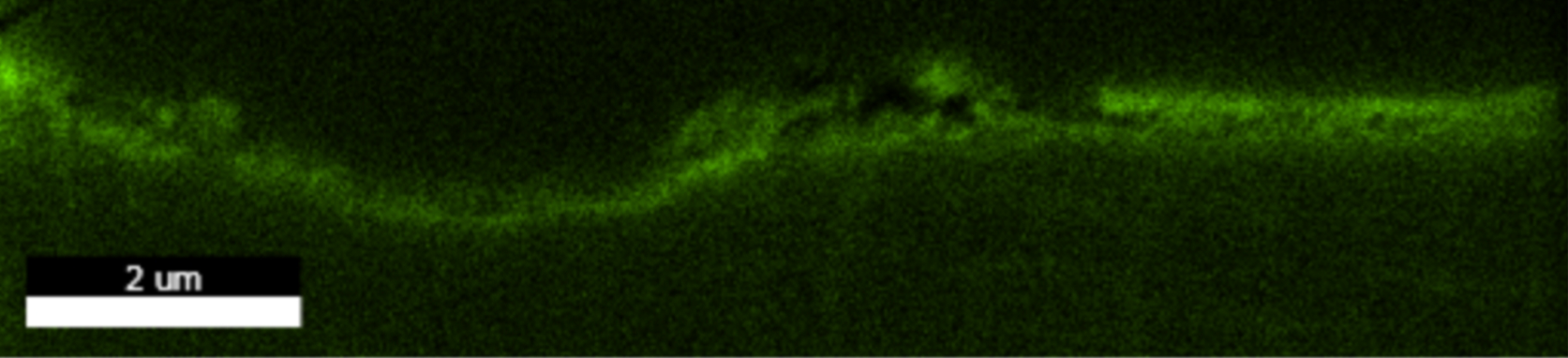}\\
  \includegraphics[width=.9\linewidth]{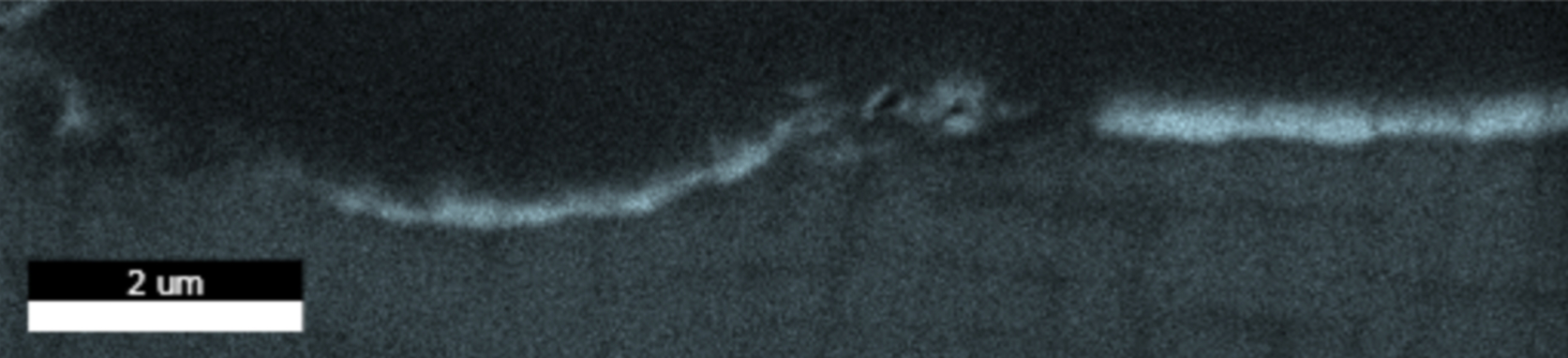}
  \caption{}
  \label{IndentCrossSection2}
\end{subfigure}
\caption{(\textbf{a}) BSE image taken at 30$^{\circ}$ during 3D cross-sectioning of an indent made at 1000$^{\circ}$C. The $\gamma - \gamma'$ microstructure is largely undistorted, with the exception of the loss of some $\gamma$ channels directly below the indent. (\textbf{b}) Oxygen (green) and aluminium (light blue) signals from an EDX map, suggesting the dark contrast in the BSE image is aluminium oxide contamination, likely from the high-temperature cements used for tip and sample fixing that formed during indentation above 700$^{\circ}$C.}
\label{IndentCrossSection}
\end{figure}

To confirm the validity of the mechanical data, a 3D cross-section of an indent made at 1000$^{\circ}$C was carried out using focussed ion beam milling to measure the residual indentation depth. Slices were cut at 20 nm intervals to ensure the deepest part of the indent could be accurately found, and the slice showing the deepest part is shown in figure \ref{IndentCrossSection} with a measured indent depth of 1.29 $\mu m$. Comparison with the hysteresis data from this indent (not shown in figure \ref{Hardness4}) shows a final depth, after the creep hold period of 1.20 $\mu m$, thereby in good agreement. The backscattered electron (BSE) image shows the $\gamma - \gamma'$ microstructure is largely undistorted, with the exception of the loss of some $\gamma$ channels directly below the indent, but there is a small amount of contamination - likely aluminium oxide - above the indent.

\subsection*{Nanoindentation Creep}
The dwell period at the peak load was used to calculate the indentation strain rate $\dot{\epsilon}$ according to equation \ref{eRate}\cite{Lucas1999601}, where $h$ is indentation depth and $t$ the time during the dwell period.

\begin{equation}
\dot{\epsilon} = \left(\frac{1}{h}\right) \left(\frac{dh}{dt}\right)
\label{eRate}
\end{equation}

The apparent stress exponent for creep, $n$, is given by equation \ref{StressExponentEq}\cite{Phani201631}, where $A$ is the uniaxial pre-exponential term. 

\begin{equation}
\dot{\epsilon} = A \sigma^n
\label{StressExponentEq}
\end{equation}

\begin{figure}[ht!]
\begin{center}
\includegraphics[width=0.66\linewidth]{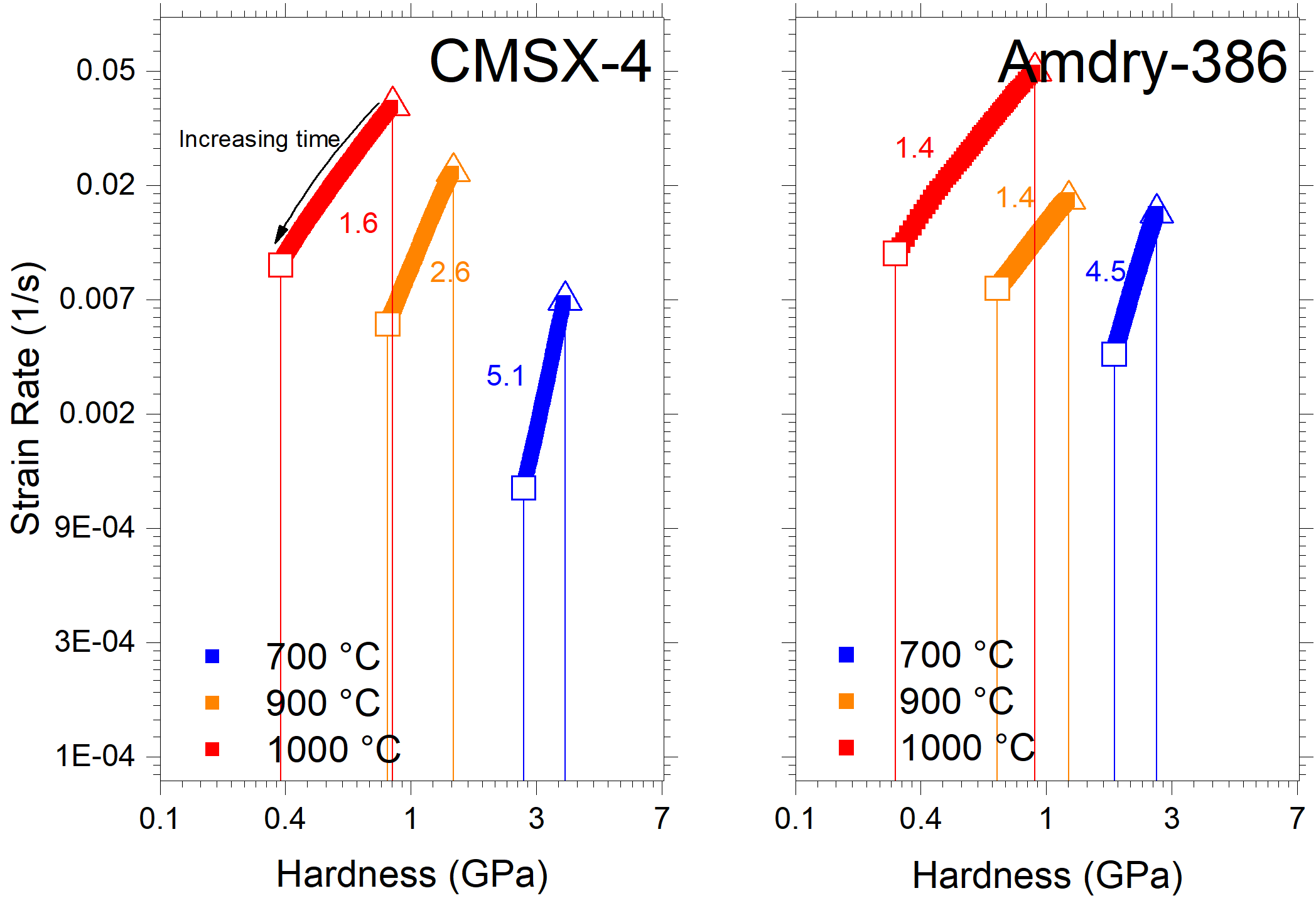}\\
\end{center}
\caption{Natural log of hardness against natural log of strain rate. The gradient at each temperature, i.e. the creep exponent, is shown. Open triangles mark the hardness at the start of the creep hold period and open squares the hardness at the end of the hold period, calculated using the instantaneous depth.\label{StressExponent}}
\end{figure}

The stress exponent was calculated using the indentation hardness in-lieu of the stress, as shown in figure \ref{StressExponent}. The indenter system used does not support a dynamic measurement of hardness throughout the hold period, therefore hardness was calculated simply as the peak load divided by the contact area from the diamond area function evaluated at the current depth to give equation \ref{StressExponentEqInd}, where $H$ is the indentation hardness.

\begin{equation}
\dot{\epsilon} = A [H(h)]^n
\label{StressExponentEqInd}
\end{equation}

As the indents were performed under a constant loading rate, these hold periods begin at differing values of strain rate. As the hold period continues, the indentation depth increases and therefore the hardness and strain rate drops. The first data point of the curve, at the beginning of the hold period, is therefore the highest and furthest to the right.

\section*{Discussion}
\subsection*{Hardness}
The nanoindentation hardness of the single-crystal CMSX-4 can be compared with the yield stress of [001]-oriented CMSX-4 as well as single-crystal $\gamma'$ ($Ni_3Al$) compression data in different crystal orientations. These data are indicated in figure \ref{Hardness} (literature data as lines for clarity, see supplemental material for full data). The hardness plateau measured here is consistent with the yield stress of CMSX-4, apart from a small size effect due to nanoindentation, also seen by other authors\cite{Sawant2008275}. The plateau, as opposed to the increasing yield stress seen in $\gamma'$, is due to the precipitation strengthening present in CMSX-4. 

The main discrepancy is the early drop in hardness: here after 400$^{\circ}$C but after 800$^{\circ}$C in the CMSX-4 yield stress and [001]-oriented compression data for Ni$_3$Al. To explain this, one must consider the different mechanisms and stress states during testing and the effect this has on the dislocation motion in the $\gamma'$ precipitates. According to the theory of Paidar, Pope, and Vitek (PPV theory)\cite{Paidar1984435}, the shear stresses acting upon the primary $(111)$ slip plane, the secondary $(1\overline{1}1)$ slip plane, and on the $\{100\}$ cube planes all influence the anomalous yielding\cite{Vitek1996135}. As a result, the peak stress is strongly dependent on crystallography. For alloys close to \textless$111$\textgreater~under tension, the peak flow stress is seen around 400$^{\circ}$C\cite{Ezz1982921}, as shown in the supplementary material, figure \ref{Literature}.

The strain underneath a Berkovich indenter extends in all directions\cite{Min200473}, allowing multiple slip planes to be stressed, and many more slip systems to operate, than in a [100] compression test. Additionally, underneath the indenter, there exist high strains (taken commonly as a representative $\sim$8\% strain\cite{FischerCripps2011}), which will lead to significant differences when comparing with bulk tension or compression tests that often use a 0.2\% proof stress to indicate yield. The strain-dependence of cross slip has been investigated by Hirsch\cite{Hirsch1992569}, who showed that the strain rate in L1$_2$ alloys is given by equation \ref{epsilon}, where $\dot{\varepsilon}$ is the strain rate, $\dot{\varepsilon}_0$ is the strain rate under zero applied stress, $k$ the Boltzmann constant, $T$ the temperature, and $H_u$ and $H_l$ are the activation enthalpies for unlocking and locking of macrokinks in the superdislocations.

\begin{equation}
\dot{\varepsilon} = \dot{\varepsilon}_0{\text{ exp}}\left(-\frac{H_u - H_l}{kT}\right)
\label{epsilon}
\end{equation}

The enthalpy of unlocking is given by equation \ref{Hu}, where ${}^0H_u$ is the enthalpy with no applied stress, $\sigma$ the applied stress and $V^{\ast}$ the activation volume. 

\begin{equation}
H_u = {}^0H_u - \sigma V^{\ast}
\label{Hu}
\end{equation}

These equations show -- somewhat intuitively -- that as the applied stress increases, dislocation motion becomes easier. The important consideration of this work is that it applies to cross-slipping superdislocations, thus explaining the observed, consistent decrease in hardness after the peak hardness at a lower temperature, as measured by nanoindentation.

To summarise: unlike uniaxial tensile data taken from yield stresses, there are high strains and a complex stress state underneath a Berkovich indenter, which in particular lead to a deviation from the zero resolved shear stress conditions on the {100} planes preventing cross-slip that occurs in uniaxial testing along [001]<001>. This therefore leads to indentation hardness measurements being strongly influenced by multiple slip systems, and cross-slip occurring in $\gamma'$ at lower temperatures which accounts for the observed drop-off in hardness. Indentation, whether at room or extreme temperatures, of the single crystal superalloy (or in fact any other strongly anisotropic material) is therefore limited in value with regards to transferability to applications where a near-uniaxial stress in a fixed direction is encountered, such as in high pressure turbine blades. However, the data presented here show that the methods employed nevertheless show good agreement with the extensive literature data available on CMSX-4.

Due to its random texture and small grain size, indentation on MCrAlY bond coats will be directly relevant for material performance in application. Although the bond coat may not carry the full centrifugal forces encountered in the blade, it is essential to the adherence of the thermal barrier coating and also used to hold abrasive particles where a shroud-design is to be avoided at the blade tip in modern engine designs. It has been shown that mechanical performance at operating temperatures is vital to retain the abrasive during operation\cite{Davenport20141877,Hardwicke2013564}. Some of the scarce literature data for MCrAlY bond coats are shown in figure \ref{Hardness}, with more detail again given in the supplementary material. The hot hardness of LCO-22 (an CoNiCrAlY alloy) also shows a similar trend to the indentation data, with the hardness dropping quickly after 400$^{\circ}$C. 

The modulus data from the two samples are also shown in figure \ref{Hardness}, along with literature data for the [001]-direction from acoustic resonance measurements on CMSX-4. The data obtained here matches that of the nanoindentation data by Sawant and Tin\cite{Sawant2008275} very closely, when their `creep correction' has not been applied. Overall, the data match closely where creep deformation is not overly significant, namely below 700$^{\circ}$C. Despite the aforementioned effects of indenter stress state, in CMSX-4 one would expect only a 10\% difference between the reduced modulus of poly-crystalline and [001]-single crystalline\cite{Vlassak19931045} samples. As the bond coat is intrinsically similar in composition, it is reasonable to find approximately the same modulus as the CMSX-4.

As previously mentioned, it is not possible to obtain reliable values of modulus from data at 900$^{\circ}$C and 1000$^{\circ}$C, where the rates of creep are significant. The hysteresis curves show significant curvature in the unloading segment, and therefore the Oliver-Pharr method is invalid. In order to obtain reliable modulus values from indentation in the case where creep is significant, the generally recommended practice\cite{ASTMNano,Chudoba2001191} is to increase the hold segment and choose the highest possible unloading rate yielding sufficient data points for the analysis. ISO 14577-1 recommends an unloading rate ten times greater than the ratio of the contact creep rate and contact compliance at maximum load. The latter may be estimated from the reduced modulus of the material from literature data (see figure \ref{Hardness}) and the diamond area function of the tip.

Considering the creep rates found here at the three highest temperatures, namely 1 nm/s at 700$^{\circ}$C, 13.5 nm/s at 900$^{\circ}$C and 16 nm/s at 1000$^{\circ}$C, target unloading rates of 19, 275 and 335 mN/s are necessary to measure the modulus. The values exceeding 100 mN are not reliably accessible in practice due to the inertia and data acquisition rate of conventional nanoindenters. Additionally, at 900$^{\circ}$C and 1000$^{\circ}$C the peak dwell segments contain no indication of the required reduction in creep rate, and dwell periods of several minutes or more might be required to achieve creep rates at which a measurement of the modulus may be feasible. Given that at high temperatures not only the accuracy of the displacement signal at long times but also sample-tip reactions are an issue, even where efficiency of the method or microstructural changes in the material need not be considered, it appears that measurement of the modulus of materials with similar creep behaviour or chemical reactivity is out of reach of nanoindentation at extreme temperatures, at least where the conventional unloading analysis is applied. Of course, a continuous stiffness measurement may improve the situation, however, this is not available for the indenter system used here.

These results therefore demonstrate an important consideration for high-temperature nanoindentation: that a single indentation test cannot reliably determine all the standard parameters, namely hardness, elastic modulus and creep resistance. These indents capture hardness and creep resistance, albeit a slightly lower hardness due to the creep hold period. However, at the highest temperatures and creep rates they are unable to capture accurate values of elastic modulus and whether measurements are likely to be successful using nanoindentation under these conditions requires careful consideration of the available methods of analysis, acquisition and unloading rates, quality of the thermal equilibrium and possible contact times between tip and sample on a case-to-case basis.

\subsection*{Creep}
\begin{figure}[ht!]
\centering
\begin{subfigure}{.5\textwidth}
  \centering
  \includegraphics[width=.9\linewidth]{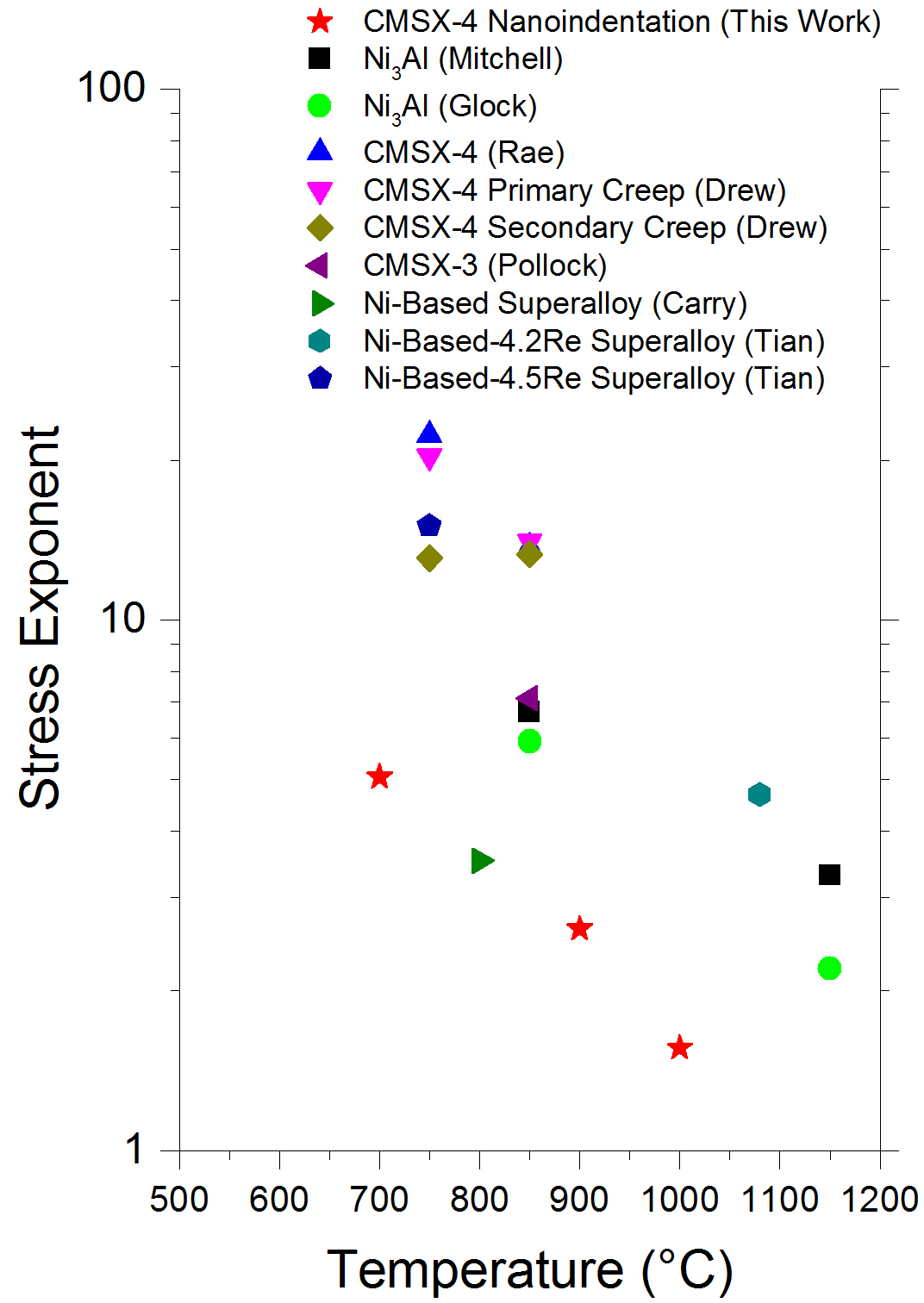}
  \caption{}
  \label{literaturestressexponent_CMSX4}
\end{subfigure}%
\begin{subfigure}{.5\textwidth}
  \centering
  \includegraphics[width=.9\linewidth]{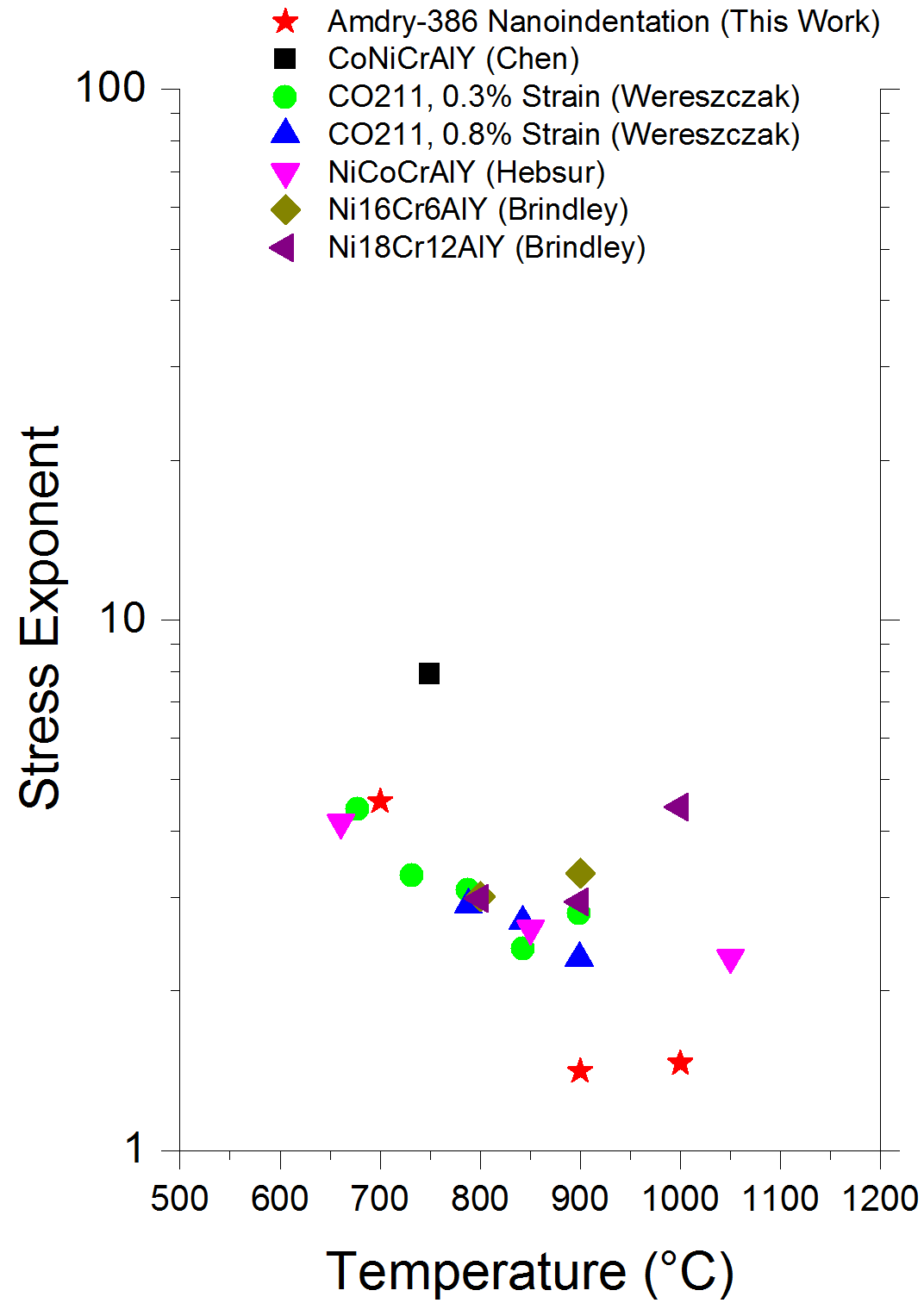}\\
  \caption{}
  \label{literaturestressexponent_Amdry}
\end{subfigure}
\caption{(\textbf{a}) A comparison of the stress exponents in CMSX-4 determined by nanoindentation and those found in the work of Mitchell\cite{Mitchell20021087}, Glock\cite{Glock20001733}, Rae\cite{Rae20071067}, Drew\cite{Drew2004127}, Pollock\cite{Pollock19921}, Carry\cite{Carry1977767}, and Tian\cite{Tian2012104,Tian2012236}. (\textbf{b}) Amdry-386 exponents are compared with the work of Hebsur\cite{Hebsur1986239,Hebsur1987143}, Wereszczak\cite{Wereszczak1998GT} (Stress relaxation under tension), Brindley\cite{Brindley199333} (Stress relaxation under tension), and Chen\cite{Chen2013205} (Punch). All tests were macroscopic tensile tests, unless otherwise stated.}
\label{LiteratureStressExponent}
\end{figure}

During the analysis of the creep data, the indentation depth was simply taken `as-is', once the hysteresis data had been corrected for thermal drift based upon the post-test thermal drift measurement. As the indent time was relatively short, the effects of thermal drift on the data are minimal and novel depth corrections used in long-term creep tests using continuous stiffness measurement\cite{Maier20151638} should not be required. Where creep is most significant (at 700$^{\circ}$C, 900$^{\circ}$C and 1000$^{\circ}$C) the determined exponents are compared with the literature in figure \ref{LiteratureStressExponent}.

A log-log plot of strain rate and hardness used to calculate the creep exponent in CMSX-4 and Amdry-386 is shown in figure \ref{StressExponent}. The 1000$^{\circ}$C curves can be seen to have a slight curvature. Kang\cite{Kang2007318} attributes a `concave' curve like this in an Mg-Sn alloy to be due to grain boundary sliding behaviour, with a `convex' curve being a result of grain matrix deformation. Similar curvatures to Kang are seen in indentation creep experiments by Mathur\cite{Mathur2016221} in the Mg$_{17}$Al$_{12}$ phase and Maier\cite{Maier20151638} in ufg-Au. Grain boundary sliding cannot be taking place in a single crystal of CMSX-4, unless significant dynamic recrystallisation is taking place. Therefore, the question is as to whether this is an effect from the sample, or one from nanoindentation. In CMSX-4, the amount of primary creep, accommodated by \textless$11\overline{2}$\textgreater$\{111\}$ slip, increases with increasing stress\cite{Reed20061}. Secondary creep then proceeds, at a time that decreases as the applied stress increases\cite{Reed20061}. Given the high stresses present, it is therefore possible that this curvature is the start of the transition to secondary creep, with the associated decrease in creep strain rate. Microstructural characterisation using TEM is needed to take place in order to confirm this. However, the curvature could simply be an artefact of the indenter velocity during rapid creep, as recent modelling results have shown\cite{Campbell20161}.

Figure \ref{LiteratureStressExponent} shows the creep stress exponents as determined by nanoindentation testing compared with values found in the literature. Apart from the data from Carry\cite{Carry1977767}, the values of stress stress exponent determined by nanoindentation in CMSX-4 are different to those found in the literature; a common problem in nanoindentation creep\cite{Goodall20065489}. These differences are often attributed to the high, changing stresses during nanoindentation creep and significant degree of primary creep. Additionally in this work, similar orientation issues to the hardness data are present: the literature is mostly concerned with [100]-oriented single crystals in tension, rather than the complex stress states present here. This may go some way to explaining the significant differences seen.

Mitchell\cite{Mitchell20021087} shows that in materials such as Ni$_3$Al with a high Peierls barrier, the apparent activation energy is strongly dependent on temperature, also seen here. Despite the aforementioned problems with the value of creep exponent, the rate of change in exponent with temperature measured by Mitchell\cite{Mitchell20021087} and Glock\cite{Glock20001733} in tension is the same as measured by indentation at 700$^{\circ}$C and above. This indicates that the nanoindentation creep is determined by the same processes as traditional creep experiments, namely double-kink nucleation to allow glide of \textless$110$\textgreater{} dislocations.

In the bond coat, the indentation exponents match very closely to those measured by tensile testing. It is not fully clear why the punch-test by Chen\cite{Chen2013205} produces significantly higher values of creep exponent, given the bond coats from Wereszczak\cite{Wereszczak1998GT} have the same composition (Co-32Ni-21Cr-8Al-0.5Y), but is likely due to the extremely complicated stress fields present in such a test. Hebsur\cite{Hebsur1986239} states that creep in the bond coat is likely a mixture of dislocation climb ($n > 5$) and viscous flow ($n = 3$), and therefore it is possible that similar creep mechanisms are occurring under both macro-scale creep tests and nanoindentation. More investigation is clearly required, but it is an encouraging first result.

\section*{Conclusions}
A CMSX-4 superalloy and its Amdry-386 bond coat were tested up to 1000$^{\circ}$C by nanoindentation. The hardness is seen to drop steadily past 400$^{\circ}$C, compared with a hardness peak at 800$^{\circ}$C as seen in single-crystal tensile tests. This differing behaviour is likely due to the complex stress field underneath a Berkovich indenter. In contrast, the hardness behaviour of the Amdry-386 is comparable to the limited literature data on other MCrAlY bond coats. The stress exponents associated with indentation creep have also been determined from the hold segments during nanoindentation. While the CMSX-4 exponents are somewhat different -- again likely due to large, complex stress fields under the indenter -- those determined in the bond coat match other MCrAlY coefficients closely.

Further work such as TEM analysis is required to examine the mechanisms of creep and deformation in the bond coat in order to determine the applicability of these data to in-service conditions. However, this work clearly demonstrates that relative comparisons of both hardness and creep resistance are able to be made on these thin layers at operationally relevant temperatures. The work thus shows that high temperature nanoindentation is well suited for future investigations of these MCrAlY coatings, in emerging new applications such as turbine sealing using abrasives\cite{Davenport20141877} or blade repair\cite{Hardwicke2013564}.

Regarding high-temperature nanoindentation, these results demonstrate the difficulty of a `one-size-fits all' approach, particular due to the influence of creep on the drift correction and modulus data. Tailored, rapid indents with no, or minimal, hold periods are ideal for the measurement of hardness, while modulus and creep information requires the use of much longer indentation time scales where possible. Finally, single crystals with strong anisotropies, such as the nickel superalloy investigated here, will show significant deviations from macroscopic investigations due to the indenter stress field and the materials to be investigated within the field of high temperature applications should therefore be selected carefully.

\section*{Additional information}
\noindent\textbf{Acknowledgements}\\
The authors would like to thank R. Va\ss en, FZJ, Germany for supplying the materials, and A. Ziemons and T. Burlet of the IMM for technical assistance. This work was supported by the Deutsche Forschungsgemeinschaft (DFG) within project KO4603/2--1.\\

\noindent\textbf{Conflict of Interest}\\
The authors declare that they have no conflict of interest.

\bibliographystyle{bibstyle}
\bibliography{1000C}

\appendix
\section{Supplementary Material}

\begin{figure}[h]
\centering
\begin{subfigure}{.5\textwidth}
  \centering
  \includegraphics[height=0.4\textheight]{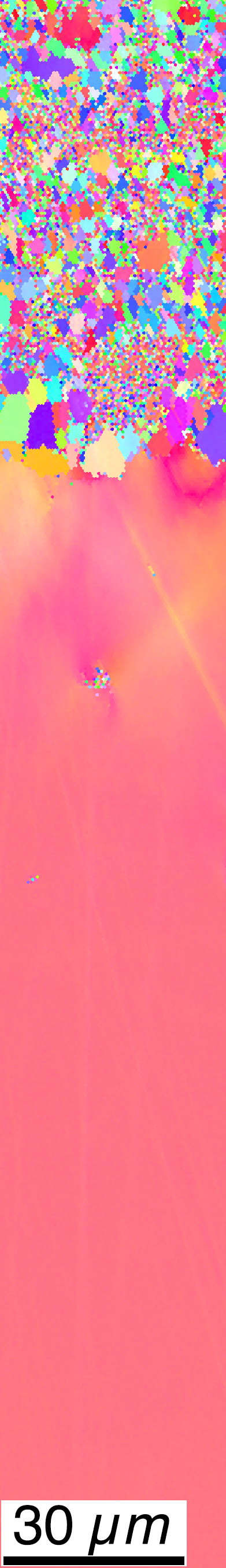}
  \caption{}
  \label{EBSDipf1}
\end{subfigure}%
\begin{subfigure}{.5\textwidth}
  \centering
  \includegraphics[width=.9\linewidth]{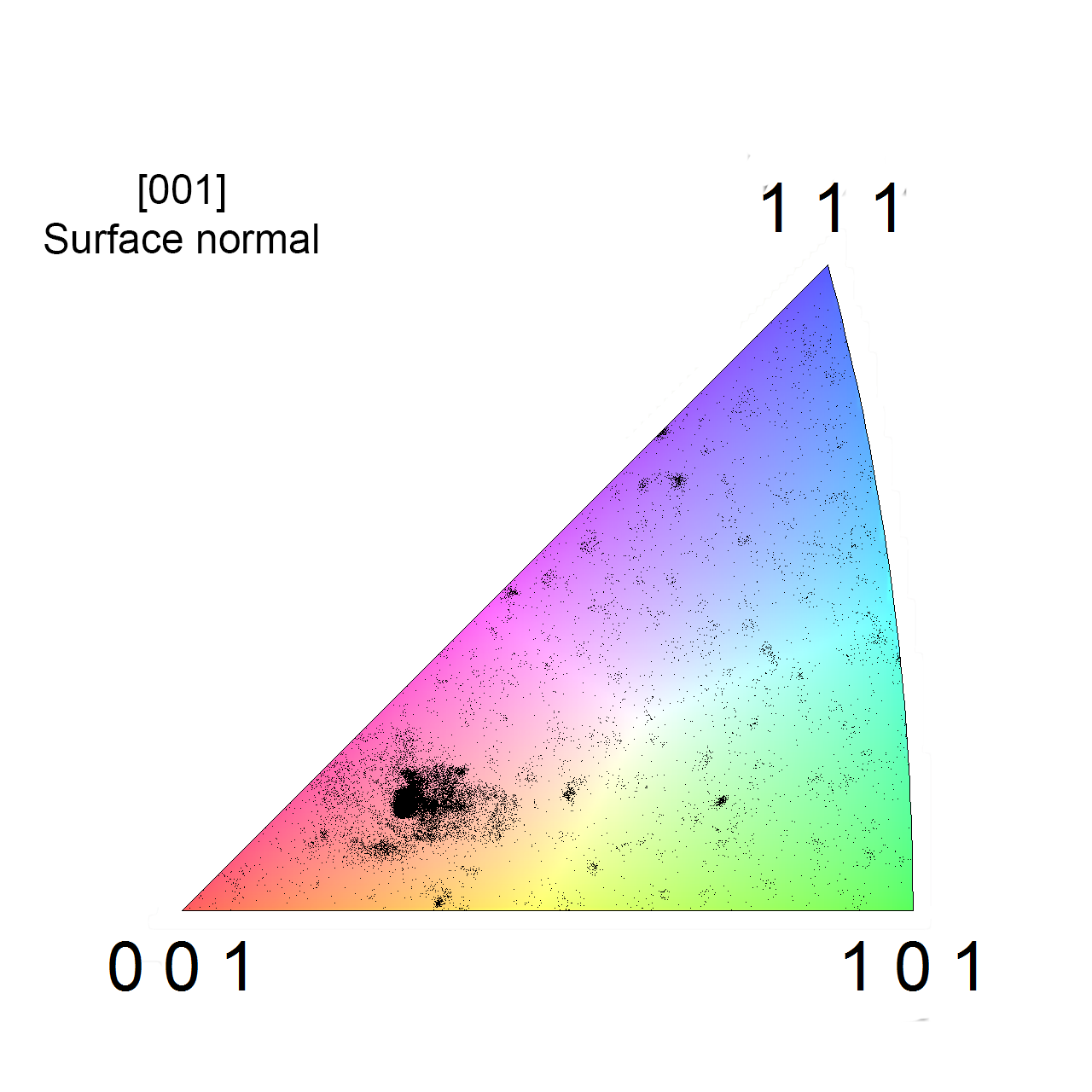}\\
  \caption{}
  \label{EBSDipf2}
\end{subfigure}
\caption{(\textbf{a}) EBSD data from a 40 $\mu m \times \text{350 } \mu m$ area encompassing some of the bond coat (top) and the bulk CMSX-4. The generated IPF for the image is also shown, (\textbf{b}). The large spot towards [100] indicates the orientation of the CMSX-4. The Amdry-386 bond coat consists of fine, equiaxed grains and shows no obvious texture.}
\label{EBSDipf}
\end{figure}

Figure \ref{EBSDipf} shows the EBSD data of some of the bond coat and the bulk CMSX-4 to which it is attached, with the IPF also shown. The CMSX-4 is close to [001], and the Amdry-386 consists of equiaxed grains a few micrometers in diameter with no overall texture.

\begin{figure}
\begin{center}
\includegraphics[width=0.8\linewidth]{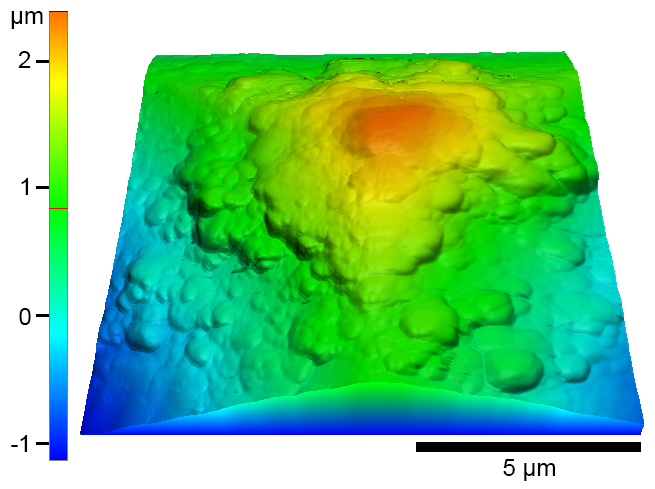}\\
\end{center}
\caption{Contact-mode AFM scan of the tip after indentation at 1000$^{\circ}$C, showing clear deposition from the sample.\label{TipAFM}}
\end{figure}

Figure \ref{TipAFM} shows an AFM scan after indentation at at 1000$^{\circ}$C, during which the measured hardness was seen to increase to over 4 GPa. As it was unlikely that this was a real effect, especially given the relatively short time spent at 1000$^{\circ}$C, the tip was scanned by AFM to determine the degree of blunting. The scan shows the loss of an accurate depth-area calibration is due to deposition on the tip. 

\begin{figure}
\begin{center}
\includegraphics[width=0.95\linewidth]{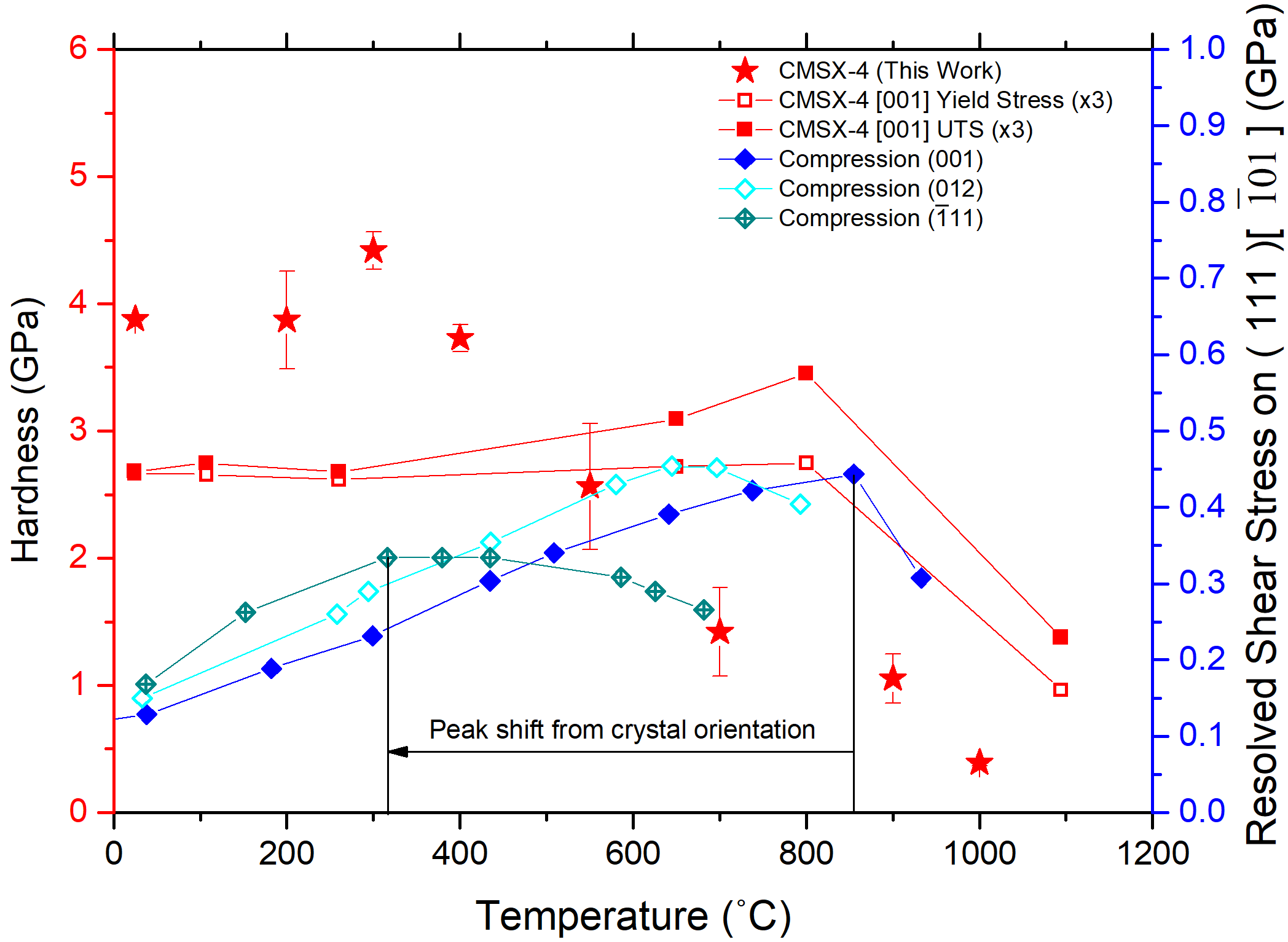}\\
\end{center}
\caption{A comparison of nanoindentation hardness with literature data: three times the yield stress and three times the UTS of [001]-oriented CMSX-4 from Sengupta\cite{Sengupta199473} and resolved shear stresses under compression of the 0.2\% off-set yield stress of various single-crystal samples of Ni$_3$Al from Ezz\cite{Ezz1982921}. \label{Literature}}
\end{figure}

\begin{figure}
\begin{center}
\includegraphics[width=0.95\linewidth]{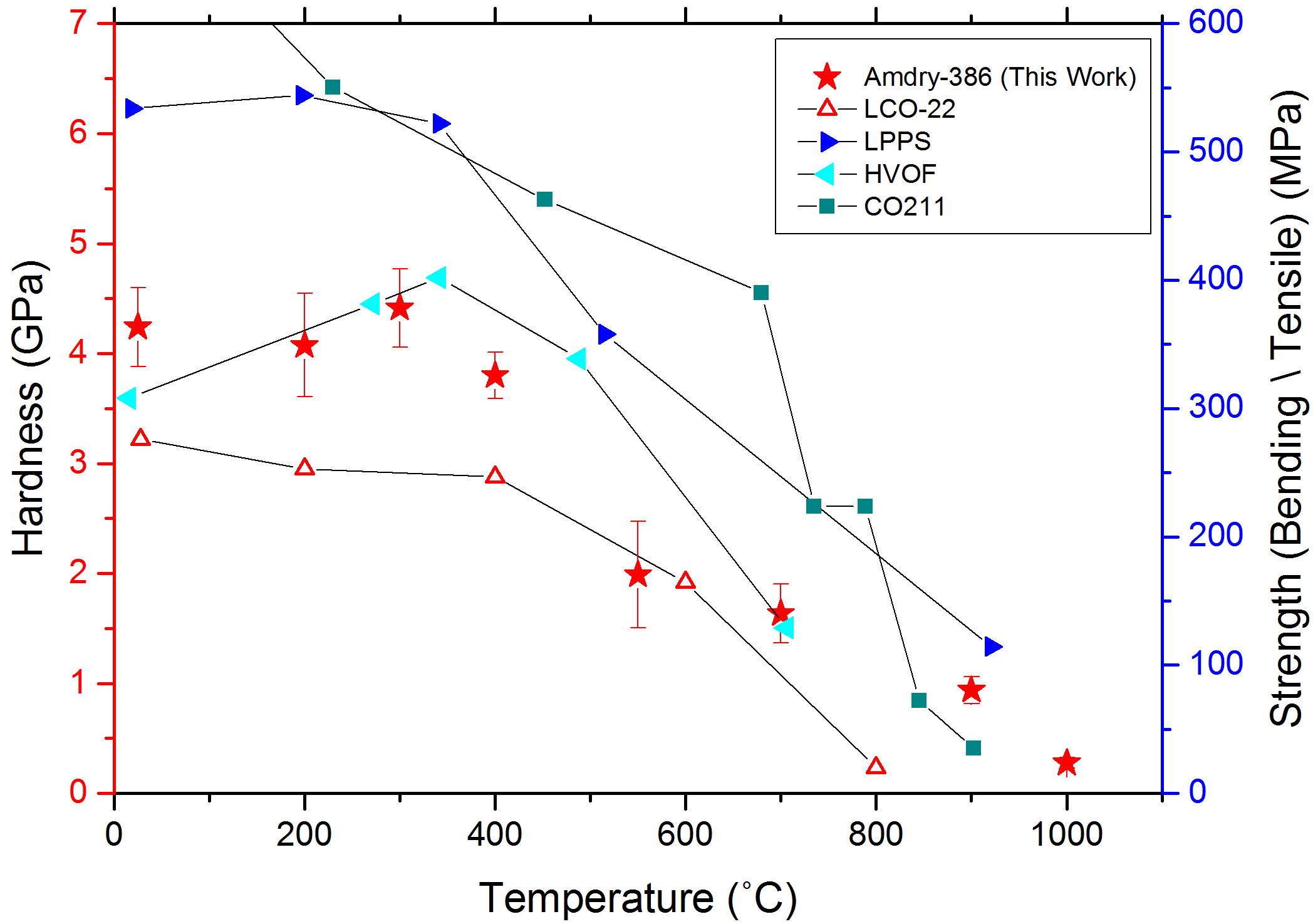}\\
\end{center}
\caption{A comparison of nanoindentation hardness with the conventionally-measured hardness of LCO-22 from Taylor\cite{Taylor19969}. The bending strengths of two different spraying techniques (LPPS and HVOF) using CoNiCrAlY coatings from Waki\cite{Waki2009500} are also shown.\label{LiteratureAmdry}}
\end{figure}

Figures \ref{Literature} and \ref{LiteratureAmdry} show in more detail a comparison of the measured indentation hardnesses with relevant literature data for the CMSX-4 and Amdry-386, respectively. 

\begin{figure}
\begin{center}
\includegraphics[width=0.8\linewidth]{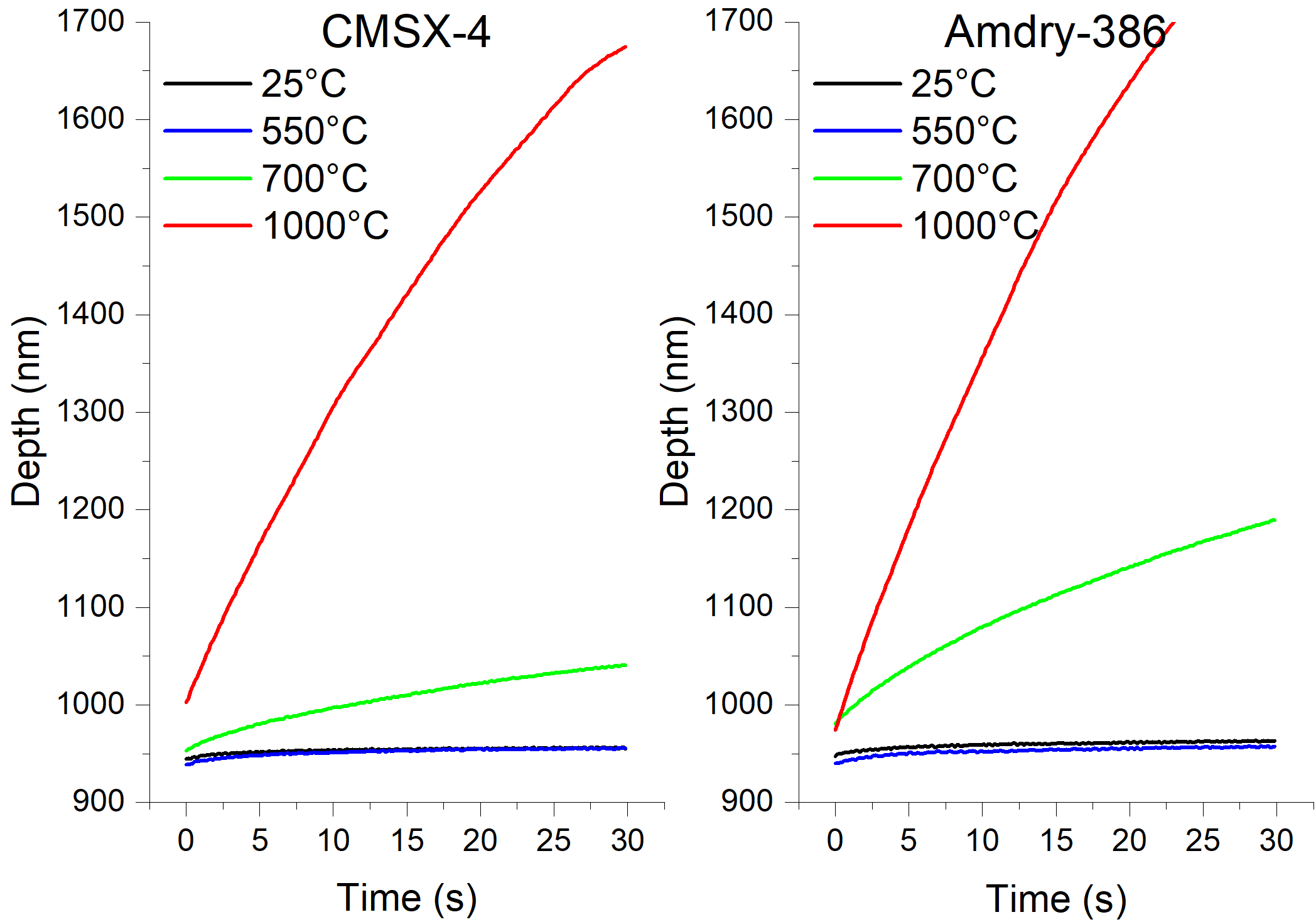}\\
\end{center}
\caption{Displacement against time for the creep hold periods are different temperatures.\label{Dwell}}
\end{figure}

Figure \ref{Dwell} shows the indenter displacement against time recorded at peak load during the creep hold period at four different temperatures. The extreme rate of displacement can be seen during the 1000$^{\circ}$C hold period due to the rapid creep rate.

\end{document}